\documentclass[a4paper]{article}

\usepackage{INTERSPEECH2022}
\usepackage[hyphens]{url}
\usepackage{hyperref}

\title{Challenges and Opportunities in Multi-device Speech Processing}
\name{Gregory Ciccarelli, Jarred Barber, Arun Nair, Israel Cohen, Tao Zhang}
\address{Amazon Alexa AI}
\email{\{gcciccar, barbjarr, aanair, isrcohen, taozhng\}@amazon.com}

\begin{document}

\maketitle
\begin{abstract}

We review current solutions and technical challenges for  automatic speech recognition,  keyword spotting, device arbitration, speech enhancement,  and  source  localization  in  multi-device  home  environments  to  provide  context  for  the  INTERSPEECH  2022 special session, ``Challenges and opportunities for signal processing and machine learning for multiple smart devices''.  We also identify the datasets needed to support these research areas. Based on the review and our research experience in the multi-device domain, we conclude with an outlook on the future evolution of multiple device signal processing and machine learning.

\end{abstract}
\noindent\textbf{Index Terms}: Speech recognition, speech enhancement, keyword spotting, arbitration.

\section{Introduction}

\nocite{pasha2020distributed}
The increasing prevalence of smart devices offers tremendous opportunities to improve user experience by leveraging spatial diversity and distributed computational and memory capability. On the other hand, local multi-device networks introduce new algorithmic and user experience challenges compared to using a single smart device, such as selecting which of several devices should perform a requested action a.k.a. device arbitration.

The core motivation of this paper is that processing on multiple devices builds upon single device research while also enabling novel algorithms and performance improvement. Robust speech recognition, enhancement, and analysis are foundational areas of speech signal processing. However, multiple devices add a new dimension- space- and a new challenge - time synchronization.

This paper contributes: (a) a signal model which is a unifying framework for understanding advances in smart device signal processing; (b) a survey of algorithmic advances in automatic speech recognition (ASR), keyword spotting (KWS), speech enhancement, and source localization; and (c) corresponding challenges and multi-device datasets for guiding future research directions. We acknowledge the detailed review published earlier by Pasha et al. \cite{pasha2020distributed}.  

In Section \ref{sec:signal_model} we define the scope of our paper and introduce a unifying signal model. In Section \ref{sec:survey}, we carry this signal model through into surveying research in automatic speech recognition, speech enhancement, and source localization. In Section \ref{sec:datasets}, we discuss the specific data needed to enable multiple device research. In Section \ref{sec:discussion}, we conclude with a discussion of an expanded view of multi-device signal processing opportunities.

\section{Signal Model}

\label{sec:signal_model}

We propose a mathematical model of multiple smart device processing to contextualize algorithms in subsequent sections. We define multiple-device signal processing as two or more distinct acoustic sensing devices with some level of onboard processing and ability to communicate with each other and/or a central server. The devices are spatially distributed on the scale of a single room or a house. We will treat extensions of this scope in the Discussion section.

We define a signal model as follows: consider a number of heterogeneous devices, indexed by $i$, in an acoustic environment (e.g., a room or a house). Each device may have a few microphones. In the environment, there are a number of localized sound sources $x_k(t)$. These sources include speech, environmental noise sources, and possibly audio being transmitted from other devices or the smart devices themselves. The acoustic impulse response between source $k$ and device $i$ is described by  $\mathbf h_i^k(t)$.  The term $\mathbf h_i^k(t)$ captures the composite impulse response of the source, room, and device. We use bold face vector notation to describe multiple channels per device.  We model the signal $\mathbf y_{i}(t)$ received at each device by:
\begin{equation}
    \mathbf y_{i}(t) = \sum_k \mathbf h_i^k(t) * x_k(t) + \mathbf n_{i}(t)
    \label{eq:signal_model}
\end{equation}
where $\mathbf n_{i}(t)$ is per-device diffuse and sensor noise. In the multiple device case, $\mathbf n_i(t)$ is statistically independent from $\mathbf n_j(t)$ ($j \neq i$).

When considering signal readouts from multiple devices, the clocks of the devices may have different sampling rates, both in gross terms (e.g. 16 kHz vs 8 kHz) and in accuracy (e.g. clock rates that differ from a nominal 16 kHz by less than 2 to more than 400 parts per million \cite{Guggenberger2015AnAO}) and also may drift over time. Furthermore, devices themselves may not at all times be active and/or in the same absolute or relative position in the room/house.
Different speech processing tasks focus on different parts of this combined signal model. In speech enhancement, we are interested in removing or mitigating the effects of noise and reverberation; that is, recovering a specific source signal $x_k(t)$ of interest. In speech recognition and keyword spotting, we want to fuse information across devices to extract information from a specific signal $x_k(t)$ while being robust to reverberation and noise. In source localization and arbitration, we are interested in properties derived from the impulse responses $\{\mathbf h_i^k(t)\}_{i}$ themselves. 

\section{Multi-device Speech Processing}
\label{sec:survey}
We describe current approaches, challenges, and future directions for several classes of speech processing tasks.
\subsection{Speech Recognition}

\label{sec:asr_kws}

Automatic speech recognition (ASR) is the translation of speech into text. This text can then be processed as instructions for a device or devices. Keyword spotting (KWS) is a special case of ASR where the goal is to detect the presence of a specific word (or small set of words). Keyword spotting is often used as the gating function for digital assistants such as Amazon’s Alexa, Google’s Assistant, and Apple’s Siri which can be activated with keywords such as “Alexa”, “Hey Google”, and “Hey Siri”. Only once the device is activated would a full ASR system be employed. Gating ASR with KWS respects user privacy and also saves power consumption.

ASR/KWS in general are difficult problems but exceptional advances have been made in the last decade with the use of deep neural networks on single devices with single or multiple microphones \cite{Hinton2012DeepNN,  Park2019SpecAugmentAS, Heymann2017BeamnetET}. These ML techniques have outperformed purely classical statistical signal processing approaches, and helped redefine signal processing to mean both classical techniques and also ML approaches. For ASR, the use of multiple distributed devices is a growing area of research with promise for further reducing word error rate.  For KWS, multiple devices may also reduce false accepts, when the device wakes up and no keyword is present, and false rejects, when the device incorrectly ignores a spoken keyword. We survey the research in this area for the various challenges and proposed current solutions.

Room reverberation as well as point source interferers and diffuse noise challenge accurate ASR in a home setting. Noise degrades signal-to-noise ratio, and environmental acoustics create complex reverberation profiles which themselves may be non-stationary if the speaker is moving through the environment. Physically distinct, spatially separate devices permit choosing high SNR streams and cancelling interferences because each device can be viewed as part of a virtual receive array of microphones in an arbitrary geometry. However, successful exploitation of multiple devices requires appropriate subselection or combination of input streams and either time alignment or robustness to time misalignment of the streams because of differing clocks and/or wireless transmission packet arrival times to a central server.

Time alignment/robustness solutions are dominated by a simple cross-correlation approach with an arbitrarily designated reference stream \cite{Yoshioka2019MeetingTU,Weninger2021DualEncoderAW}. While there is nothing reproachable about a simple, robust algorithm, time alignment is potentially the first challenge that confronts a multi-device system and is pervasive in any multi-device setting. Consequently, despite existing work on time alignment and beamforming, we see a lack of studies regarding the robustness of end to end algorithms that build on potentially misaligned streams as an important area of study.

Source selection and fusion solutions are varied, and they can take place at the level of selecting or fusing all waveforms from one device from multiple devices or selecting particular waveforms from several devices each of which may have multiple microphones. Initial solutions included computing traditional interpretable speech features of signal quality on each individual stream (raw microphone waveform or beamformed output from a device), and using a heuristic assignment of weights to compute either a weighted sum, or ranking of which stream was the most reliable \cite{BentezGuijarro2019CoordinationOS}. Since then, more sophisticated and/or optimal approaches that may also account for inter-device communication cost have been proposed \cite{araki2018comparison, zhang2020study, casebeer2021communication}.   Weninger et al.~\cite{Weninger2021DualEncoderAW} used an ML approach with an encoder selector network to learn a decision making front end that weights between close talk and far talk streams. In addition to stream fusion is stream posterior fusion. ROVER \cite{1997fiscus} was an early, successful voting  approach to fusion, and has been followed by Confusion Network Combinations \cite{Evermann2000PosteriorPD}.


\subsection{Speech Enhancement and Interference Cancellation}

\label{sec:speech_enhancement}

Speech enhancement \cite{loizou2007speech} is the process of improving the quality or intelligibility of speech within an audio stream. Speech enhancement is often a pre-processing step in applications spanning human listening use cases (VoIP, hearing aids) to machine processing tasks (ASR, KWS). With the continued proliferation of mobile, wearable, and smart-home audio capture devices, the challenge of speech enhancement has grown - solutions are now needed for increased speaker-microphone distances, diverse acoustic and noise environments, and limited compute and storage budgets.

To tackle more difficult acoustic environments, audio capture devices transitioned to having multiple microphones and speech enhancement research focused on methods to exploit the signal diversity inherent to these multichannel interfaces \cite{gannot2017consolidated} to improve performance. Concurrent with advancements in multichannel microphone array processing, advances in machine learning have been critical to achieving current speech enhancement and recognition performance. Systems expertly blending classical signal processing with deep neural networks \cite{haeb2019speech} are at the core of the smart home devices with which we populate our homes.

A family of speech enhancement approaches extends ML DNN methods designed for the single device multichannel case to the multi-device multichannel case. This family can be subdivided into two categories - the first category consists of methods that learn single-channel time-frequency masks to apply to individual channels before performing data-dependent beamforming - either Generalized Eigenvalue (GEV) beamforming \cite{heymann2016neural} or Minimum-Variance Distortionless Response (MVDR) beamforming \cite{higuchi2016robust}. This category of methods generalizes to unknown distributed microphone array geometries as both components - the mask estimation and the data-driven beamforming - can be flexibly applied on any array configuration. However their strength is also a drawback. These approaches solely focus on learning single-channel masks, and consequently they ignore the spatial inter-relationships between channels. The second category of more recent approaches seeks to overcome this drawback by explicitly learning spatial relationships directly by taking multichannel audio data or cross-channel features as input to the DNN and performing end-to-end speech enhancement. These neural beamformers started off being designed for the single device multichannel case where microphone array geometry is known \emph{a priori} and the same across train and test scenarios \cite{meng2017deep, tolooshams2020channel}. Recent breakthroughs have allowed for array geometry-agnostic multichannel deep neural beamformers \cite{zhang2021microphone, taherian2021one, pandey2021tadrn}

Another family of speech enhancement approaches explicitly considers the local processing scenario where speech processing is done on-device a.k.a. at the edge. The first category here consists of methods where one device (dubbed the “fusion center” \cite{wang2019distributed} ) has significantly more processing power than the other devices. In these scenarios, each recording device transmits recorded audio to the hub which runs the speech enhancement algorithm \cite{ali2021integrated}. The second category of approaches distribute the processing over the whole set of devices in order to remove the need for a fusion center while still keeping inter-node communication bandwidth considerations in mind \cite{furnon2021dnn}.

A related field important to improving speech quality is Acoustic Echo Cancellation (AEC). In the single device case, AEC allows the same device to simultaneously both play audio and listen to voice commands by using a reference copy of the clean playback signal to suppress the interfering playback signal in the microphone signals without distorting the target speech. AEC allows devices to operate in challenging regimes ($\leq$ -30 dB Signal-to-Echo Ratios) while delivering better performance than reference-agnostic noise cancellation.

However, generalizing AEC to the multi-device case, where devices might be networked wirelessly, e.g., through WiFi or Bluetooth, introduces the additional challenge of time-varying delays as a result of imperfect channel synchronization. This coupled with a lack of clock synchronization across devices hinders performance. To combat these issues, Ayrapetian et al.~\cite{ayrapetian2021asynchronous} proposed using a combination of multichannel AEC together with beamforming to achieve acoustic echo suppression in excess of 40 dB.

In terms of performance, speech enhancement is usually quantified with Perceptual Evaluation of Speech Quality (PESQ) \cite{rix2001perceptual}, Short-Time Objective Intelligibility (STOI) \cite{taal2010short}, Signal-to-Distortion Ratio (SDR) \cite{le2019sdr}, or Word Error Rate (WER) on a downstream speech recognition system. AEC is usually measured in terms of Echo Return Loss Enhancement (ERLE) or performance on a downstream task e.g., False Rejection Rate (FRR) in KWS systems.

While providing absolute performance numbers is difficult as different methods use different datasets and are specialized to different scenarios, we quote the performance numbers from a recent work \cite{taherian2021one} trained and tested on data from the DNS-Challenge dataset \cite{reddy2021interspeech}. A microphone array geometry agnostic model tested on an unseen array geometry obtained a WER of 20.22, SDR of 12.51, and STOI of 88.45 (See Table 3, Dataset A, Geometry Agnostic result for the Circular (5ch) microphone array) compared to a WER of 24.00, SDR 11.41, and STOI of 85.77 for the baseline Signal Averaging model that processed each microphone stream independently with a single-channel speech enhancement model followed by averaging the enhanced signals.  This results showcased the benefit of joint stream processing.

Multi-device distributed speech enhancement and echo cancellation is still a nascent field. Taherian et al.~\cite{taherian2021one} observe that spatial aliasing is a known issue and an open research question. In addition, distributed speech enhancement algorithms designed to explicitly trade off performance with inter-device communication costs and latency requirements need additional research. For AEC, explicit clock compensation and channel synchronization on device promises better performance if compute cost barriers can be overcome.  Finally, the parallels between ASR/KWS and speech enhancement are prompting researchers to consider jointly solving both problems simultaneously.

\subsection{Source Localization}

The source localization problem, including variants such as direction-of-arrival (DOA) estimation and device arbitration, is to extract spatial information relationships between sources and devices. In contrast to other tasks, we are interested in information contained in the impulse responses themselves, rather than the signals; the role of the signals $x_k(t)$ and the impulse responses $\mathbf h_i^k(t)$ are reversed relative to in ASR/KWS and speech enhancement. 

In the multiple smart device case, understanding the spatial relationships between a speaker and a number of devices may improve how a system understands the intent of and interacts with the user.

A special case of source localization is \emph{device arbitration} \cite{barber2021endtoend}. In device arbitration, we move from a regression problem to a classification problem: we are interested in identifying the \emph{closest} device to the speaker, without concern for physical ranges. This simplification can be exploited to achieve better system performance than performing independent localization of each device to the source.

Some classical methods for localization, such as time difference of arrival (TDOA), require accurate clock synchronization between devices \cite{locreview}. This can be especially challenging in ad-hoc arrays of consumer-grade devices, where unsynchronized clocks and cross-device communication latency must be considered. Some authors have proposed fusing audio and visual features in a Simultaneous Localization and Mapping (SLAM) framework to jointly localize devices and speakers \cite{michaud20203d}.  Barber et al. \cite{barber2021endtoend} has innovated by introducing an end to end ML approach to device arbitration.  Because we typically do not know the locations of devices relative to each other in a user's environment, these algorithms also are hindered by a lack of effective on-line evaluation.


\section{Multiple Device Datasets}

\label{sec:datasets}

To support multi-device signal processing and ML research summarized above, multi-device datasets either real or simulated are needed. While a purely simulated dataset has the advantage of minimal cost, room geometries may necessarily be simplified to ease computational complexity. Therefore, measured responses are the gold standard in the sense of capturing real-world physical interactions. However, such measurements can be costly both in time and labor.  To strike a balance between realism and time/cost, a common approach is to measure room impulse responses and then to invoke an assumption of linearity such that dry speech (speech collected in an anechoic room) is convolved with the real, measured room impulse response (RIR) to create the speech (or more generally, sound) sample used for research. Returning to our signal model, the objective is to reliably measure or create the impulse responses $\mathbf h_i^k(t)$ between a source $k$ and the microphones on a device $i$. 


Multiple device data collections are actually not as common as one might first assume from searching for room impulse response databases. We require that the source position be fixed and multiple receive positions be known. Having one source and N simultaneous receivers is a natural setup. However, as long as the source itself is not moved, and we assume that the presence of the devices themselves has minimal impact on the RIRs between other devices and the source, we can consider “virtual collections” of N sources. The idea of a virtual collection of N devices is especially advantageous for scenarios of device arbitration. As device arbitration, at its most simple, is to choose between two devices relative to a single source, N devices yields N choose 2 arbitration scenarios. Increasing N yields a combinatorial explosion which is favorable in that many, many more scenarios can then be tested.

We also recognize that different areas of research impose more stringent standards on the metadata and precision of the spatial configurations and room geometry/characterization. Home theater applications are the most stringent because of the sensitivity of the human ear. Absolute 6 degrees of freedom characterization of the relative positions of the devices is required. Device arbitration is the next most stringent and further imposes the requirement of “broad” spatial diversity- device arbitration is not as often concerned with choosing between two devices close to each other in the same room, but in multiple rooms that are acoustically connected. ASR, KWS, speech enhancement, and de-reverberation generally do not need positional information at all as their aim is to determine what acoustic event took place accurately in a realistic environment, not to locate it.

Table~\ref{tbl:datasets} lists data sets that met our strict inclusion criteria of being friendly for commercial research, being easily downloaded, and supporting either virtual collections of devices or directly measuring simultaneously the RIR to different receiver locations. Other tasks such as virtual reality simulation with binaural responses and  reverse engineering the room dimensions were outside of the scope of our survey.

\begin{table}
\centering
\caption{Open Source RIR Datasets}
\begin{tabular}{p{3cm} p{4cm}}
\hline
Dataset & Contents\\
\midrule
BUT Speech at FIT Reverb Database \cite{2019szoke} & $\sim$1400 RIRs from 31 simultaneous microphones across 9 rooms  \\
Transition between rooms \cite{McKenzie2021AcousticAA} & $\sim$400 RIRs sampled along transition path between four pairs of rooms  \\
Spatial room impulse responses \cite{McKenzie2021DatasetOS} & 100 RIRs in a varachoic room  \\
ACE Challenge \cite{Eaton2016EstimationOR} & 70 RIRS from 5 simultaneous mic arrays across 7 rooms, 2 positions per room, plus 58 anechoic speech utterances.\\
dEchorate \cite{di2021dechorate} & 1800 RIRs from 30 simultaneous mics in a varachoic room\\
\hline
\end{tabular}
\label{tbl:datasets}
\end{table}

\section{Discussion \& Future Directions}

\label{sec:discussion}

We have covered progress in ASR/KWS, speech enhancement, and device arbitration as well as existing datasets for multi-device research. We can situate the goals of these research areas in a multi-device signal model understanding of the problem space.  Multi-device processing  distinguishes itself from single device processing in that developers must coordinate distributed devices in virtual arrays with absent or poor clock synchronization, unknown relative microphone positions, and differences in gains on devices to capitalize on the  additional degrees of sensing in both space and time  \cite{pasha2020distributed}.


From our review, the following set of demands on the talents of researchers have emerged: (a) the design of ASR/KWS and speech enhancement algorithms that are accurate and robust in diverse, non-stationary, noisy and reverberant acoustic environments while also being highly energy efficient as edge based computing is important for user privacy but also runs constantly as a background process; (b) the development of device arbitration algorithms that are consistent with user expectations and robust against diverse acoustic environments; (c) the construction of multi-device datasets to support the above multi-device signal processing and ML research.


Beyond the acoustic signal processing and ML directions we have highlighted, we foresee an evolution in the definition of multi-device processing. Specifically, smart homes and smart environments will be comprised of ad-hoc networks in which the processing is not siloed within a single entity's operating system stack. This diffusion of processing control is something that will only become more important as the number of organizations innovating for users increases. Lippi et al.~\cite{2017lippi} describe one vision of this future in which another layer of communication and processing protocols needs to be established between devices and also perhaps in conjunction with an information broker.

Media induced false wake in KWS is a concrete example of a present need for communication between currently unlinked devices. Mass media presentation of a keyword that is picked up by hundreds or thousands of devices simultaneously is a pain point for users and developers today. Consequently, engineering systems from different organizations to communicate relevant information to avoid system breakdown and frustration with both organizations is a crucial area of future work.

We have primarily discussed acoustic signal processing with multiple devices, but in addition to the proliferation of devices is also the proliferation of sensing capabilities and modalities on those devices. Early proof of promise in this direction includes audio-visual speech recognition \cite{2015mroueh}. In the spirit of multiple device processing, there will also be non-traditional signal acquisition streams. Biotechnology and physiological sensing is a prime example of a rich space that as yet has little cross pollination with audio visual fields, though we do see paralinguistic research in speech making inroads in health monitoring applications.

Taking a step farther back, we challenge practitioners to think really big- what does multiple device signal processing mean at the scale of millions or billions of connected devices? These include not just the in-home devices, but ``on the go" devices such as our automobiles, fitness trackers, and smartphones. A glimpse of the potential of these mega-networks has been realized with the fusion of data for COVID-19 tracking. What additional use cases in public health, transportation planning, and resource allocation decisions are untapped?

As exciting as the idea of a future of communicating mega networks might be to some, we want to temper this enthusiasm with the importance of keeping and maintaining user trust in these systems. Trust is the bedrock upon which any technology must be situated. What information is captured, retained by whom and for how long, and communicated to other parties is crucial at every scale of multi-device signal processing- from two smart sensors in a home to millions of devices from multiple organizations around the world. We are optimistic that algorithmic breakthroughs in multi-device signal processing and ML such as federated learning will enable safeguards for user privacy and security while also delivering enhanced performance and value.


\bibliographystyle{IEEEtran}
\bibliography{mybib}

\begin{thebibliography}{10}
\providecommand{\url}[1]{#1}
\csname url@samestyle\endcsname
\providecommand{\newblock}{\relax}
\providecommand{\bibinfo}[2]{#2}
\providecommand{\BIBentrySTDinterwordspacing}{\spaceskip=0pt\relax}
\providecommand{\BIBentryALTinterwordstretchfactor}{4}
\providecommand{\BIBentryALTinterwordspacing}{\spaceskip=\fontdimen2\font plus
\BIBentryALTinterwordstretchfactor\fontdimen3\font minus
  \fontdimen4\font\relax}
\providecommand{\BIBforeignlanguage}[2]{{%
\expandafter\ifx\csname l@#1\endcsname\relax
\typeout{** WARNING: IEEEtran.bst: No hyphenation pattern has been}%
\typeout{** loaded for the language `#1'. Using the pattern for}%
\typeout{** the default language instead.}%
\else
\language=\csname l@#1\endcsname
\fi
#2}}
\providecommand{\BIBdecl}{\relax}
\BIBdecl

\bibitem{pasha2020distributed}
S.~Pasha, J.~Lundgren, C.~Ritz, and Y.~Zou, ``Distributed microphone arrays,
  emerging speech and audio signal processing platforms: A review,'' \emph{Adv.
  in Sci., Tech. and Eng. Sys.}, vol.~5, no.~4, pp. 331--343, 2020.

\bibitem{Guggenberger2015AnAO}
M.~Guggenberger, M.~Lux, and L.~B{\"o}sz{\"o}rm{\'e}nyi, ``An analysis of time
  drift in hand-held recording devices,'' in \emph{MMM}, 2015.

\bibitem{Hinton2012DeepNN}
G.~E. Hinton, L.~Deng, D.~Yu, G.~E. Dahl, A.~Mohamed, N.~Jaitly, A.~W. Senior,
  V.~Vanhoucke, P.~Nguyen, T.~N. Sainath, and B.~Kingsbury, ``Deep neural
  networks for acoustic modeling in speech recognition,'' \emph{IEEE Sig. Proc.
  Mag.}, vol.~29, p.~82, 2012.

\bibitem{Park2019SpecAugmentAS}
D.~S. Park, W.~Chan, Y.~Zhang, C.-C. Chiu, B.~Zoph, E.~D. Cubuk, and Q.~V. Le,
  ``Specaugment: A simple data augmentation method for automatic speech
  recognition,'' in \emph{INTERSPEECH}, 2019.

\bibitem{Heymann2017BeamnetET}
J.~Heymann, L.~Drude, C.~B{\"o}ddeker, P.~Hanebrink, and R.~H{\"a}b-Umbach,
  ``Beamnet: End-to-end training of a beamformer-supported multi-channel asr
  system,'' \emph{{}ICASSP}.

\bibitem{Yoshioka2019MeetingTU}
T.~Yoshioka, D.~Dimitriadis, A.~Stolcke, W.~Hinthorn, Z.~Chen, M.~Zeng, and
  X.~Huang, ``Meeting transcription using asynchronous distant microphones,''
  in \emph{INTERSPEECH}, 2019.

\bibitem{Weninger2021DualEncoderAW}
F.~Weninger, M.~Gaudesi, R.~Leibold, R.~Gemello, and P.~Zhan, ``Dual-encoder
  architecture with encoder selection for joint close-talk and far-talk speech
  recognition,'' \emph{IEEE ASR and Understanding Workshop}, pp. 534--540,
  2021.

\bibitem{BentezGuijarro2019CoordinationOS}
A.~Ben{\'i}tez-Guijarro, Z.~C. Carri{\'o}n, M.~Noguera, and K.~B. Akhlaki,
  ``Coordination of speech recognition devices in intelligent environments with
  multiple responsive devices,'' in \emph{UCAmI}, 2019.

\bibitem{araki2018comparison}
S.~Araki, N.~Ono, K.~Kinoshita, and M.~Delcroix, ``Comparison of reference
  microphone selection algorithms for distributed microphone array based speech
  enhancement in meeting recognition scenarios,'' in \emph{Intl. Wk. on
  Acoustic Sig. Enh.}\hskip 1em plus 0.5em minus 0.4em\relax IEEE, 2018, pp.
  316--320.

\bibitem{zhang2020study}
J.~Zhang, H.~Chen, L.-R. Dai, and R.~C. Hendriks, ``A study on reference
  microphone selection for multi-microphone speech enhancement,''
  \emph{IEEE/ACM Trans. on Audio, Speech, and Lang. Proc.}, vol.~29, pp.
  671--683, 2020.

\bibitem{casebeer2021communication}
J.~Casebeer, J.~Kaikaus, and P.~Smaragdis, ``Communication-cost aware
  microphone selection for neural speech enhancement with ad-hoc microphone
  arrays,'' in \emph{ICASSP}.\hskip 1em plus 0.5em minus 0.4em\relax IEEE,
  2021, pp. 8438--8442.

\bibitem{1997fiscus}
J.~Fiscus, ``A post-processing system to yield reduced word error rates:
  Recognizer output voting error reduction ({ROVER}),'' in \emph{1997 IEEE
  Workshop on Automatic Speech Recognition and Understanding Proceedings},
  1997, pp. 347--354.

\bibitem{Evermann2000PosteriorPD}
G.~Evermann and P.~C. Woodland, ``Posterior probability decoding, confidence
  estimation and system combination,'' in \emph{Proc. Speech Transcription
  Workshop}, vol.~27.\hskip 1em plus 0.5em minus 0.4em\relax Citeseer, 2000,
  pp. 78--81.

\bibitem{loizou2007speech}
P.~C. Loizou, \emph{Speech enhancement: theory and practice}.\hskip 1em plus
  0.5em minus 0.4em\relax CRC press, 2007.

\bibitem{gannot2017consolidated}
S.~Gannot, E.~Vincent, S.~Markovich-Golan, and A.~Ozerov, ``A consolidated
  perspective on multimicrophone speech enhancement and source separation,''
  \emph{IEEE/ACM Trans. Audio, Speech, and Lang. Proc.}, vol.~25, no.~4, pp.
  692--730, 2017.

\bibitem{haeb2019speech}
R.~Haeb-Umbach, S.~Watanabe, T.~Nakatani, M.~Bacchiani, B.~Hoffmeister, M.~L.
  Seltzer, H.~Zen, and M.~Souden, ``Speech processing for digital home
  assistants: Combining signal processing with deep-learning techniques,''
  \emph{IEEE Sig. Proc. Mag.}, vol.~36, no.~6, pp. 111--124, 2019.

\bibitem{heymann2016neural}
J.~Heymann, L.~Drude, and R.~Haeb-Umbach, ``Neural network based spectral mask
  estimation for acoustic beamforming,'' in \emph{ICASSP}.\hskip 1em plus 0.5em
  minus 0.4em\relax IEEE, 2016, pp. 196--200.

\bibitem{higuchi2016robust}
T.~Higuchi, N.~Ito, T.~Yoshioka, and T.~Nakatani, ``Robust {MVDR} beamforming
  using time-frequency masks for online/offline asr in noise,'' in
  \emph{{ICASSP}}.\hskip 1em plus 0.5em minus 0.4em\relax IEEE, 2016, pp.
  5210--5214.

\bibitem{meng2017deep}
Z.~Meng, S.~Watanabe, J.~R. Hershey, and H.~Erdogan, ``Deep long short-term
  memory adaptive beamforming networks for multichannel robust speech
  recognition,'' in \emph{{ICASSP}}.\hskip 1em plus 0.5em minus 0.4em\relax
  IEEE, 2017, pp. 271--275.

\bibitem{tolooshams2020channel}
B.~Tolooshams, R.~Giri, A.~H. Song, U.~Isik, and A.~Krishnaswamy,
  ``Channel-attention dense {U}-net for multichannel speech enhancement,'' in
  \emph{{ICASSP}}.\hskip 1em plus 0.5em minus 0.4em\relax IEEE, 2020, pp.
  836--840.

\bibitem{zhang2021microphone}
S.~Zhang and X.~Li, ``Microphone array generalization for multichannel
  narrowband deep speech enhancement,'' \emph{arXiv preprint arXiv:2107.12601},
  2021.

\bibitem{taherian2021one}
H.~Taherian, S.~E. Eskimez, T.~Yoshioka, H.~Wang, Z.~Chen, and X.~Huang, ``One
  model to enhance them all: array geometry agnostic multi-channel personalized
  speech enhancement,'' \emph{arXiv preprint arXiv:2110.10330}, 2021.

\bibitem{pandey2021tadrn}
A.~Pandey, B.~Xu, A.~Kumar, J.~Donley, P.~Calamia, and D.~Wang, ``{TADRN}:
  Triple-attentive dual-recurrent network for ad-hoc array multichannel speech
  enhancement,'' \emph{arXiv preprint arXiv:2110.11844}, 2021.

\bibitem{wang2019distributed}
S.-S. Wang, Y.-Y. Liang, J.-w. Hung, Y.~Tsao, H.-M. Wang, and S.-H. Fang,
  ``Distributed microphone speech enhancement based on deep learning,''
  \emph{arXiv preprint arXiv:1911.08153}, 2019.

\bibitem{ali2021integrated}
R.~Ali, T.~van Waterschoot, and M.~Moonen, ``An integrated {MVDR} beamformer
  for speech enhancement using a local microphone array and external
  microphones,'' \emph{EURASIP J. on Audio, Speech, and Music Proc.}, vol.
  2021, no.~1, pp. 1--20, 2021.

\bibitem{furnon2021dnn}
N.~Furnon, R.~Serizel, S.~Essid, and I.~Illina, ``{DNN}-based mask estimation
  for distributed speech enhancement in spatially unconstrained microphone
  arrays,'' \emph{IEEE/ACM Trans. on Audio, Speech, and Lang. Proc.}, vol.~29,
  pp. 2310--2323, 2021.

\bibitem{ayrapetian2021asynchronous}
R.~Ayrapetian, P.~Hilmes, M.~Mansour, T.~Kristjansson, and C.~Murgia,
  ``Asynchronous acoustic echo cancellation over wireless channels,'' in
  \emph{{ICASSP}}.\hskip 1em plus 0.5em minus 0.4em\relax IEEE, 2021, pp.
  116--120.

\bibitem{rix2001perceptual}
A.~W. Rix, J.~G. Beerends, M.~P. Hollier, and A.~P. Hekstra, ``Perceptual
  evaluation of speech quality ({PESQ})-a new method for speech quality
  assessment of telephone networks and codecs,'' in \emph{{ICASSP}},
  vol.~2.\hskip 1em plus 0.5em minus 0.4em\relax IEEE, 2001, pp. 749--752.

\bibitem{taal2010short}
C.~H. Taal, R.~C. Hendriks, R.~Heusdens, and J.~Jensen, ``A short-time
  objective intelligibility measure for time-frequency weighted noisy speech,''
  in \emph{{ICASSP}}.\hskip 1em plus 0.5em minus 0.4em\relax IEEE, 2010, pp.
  4214--4217.

\bibitem{le2019sdr}
J.~Le~Roux, S.~Wisdom, H.~Erdogan, and J.~R. Hershey, ``{SDR}--half-baked or
  well done?'' in \emph{{ICASSP}}.\hskip 1em plus 0.5em minus 0.4em\relax IEEE,
  2019, pp. 626--630.

\bibitem{reddy2021interspeech}
C.~K. Reddy, H.~Dubey, K.~Koishida, A.~Nair, V.~Gopal, R.~Cutler, S.~Braun,
  H.~Gamper, R.~Aichner, and S.~Srinivasan, ``Interspeech 2021 deep noise
  suppression challenge,'' \emph{arXiv preprint arXiv:2101.01902}, 2021.

\bibitem{barber2021endtoend}
J.~Barber, Y.~Fan, and T.~Zhang, ``End-to-end {A}lexa device arbitration,'' in
  \emph{{ICASSP}}.\hskip 1em plus 0.5em minus 0.4em\relax IEEE, 2022.

\bibitem{locreview}
\BIBentryALTinterwordspacing
M.~U. Liaquat, H.~S. Munawar, A.~Rahman, Z.~Qadir, A.~Z. Kouzani, and M.~A.~P.
  Mahmud, ``Localization of sound sources: A systematic review,''
  \emph{Energies}, vol.~14, no.~13, 2021. [Online]. Available:
  \url{https://www.mdpi.com/1996-1073/14/13/3910}
\BIBentrySTDinterwordspacing

\bibitem{michaud20203d}
S.~Michaud, S.~Faucher, F.~Grondin, J.-S. Lauzon, M.~Labb{\'e},
  D.~L{\'e}tourneau, F.~Ferland, and F.~Michaud, ``3{D} localization of a sound
  source using mobile microphone arrays referenced by {SLAM},'' in
  \emph{IEEE/RSJ Intl. Conf. on Intell. Robots and Sys.}\hskip 1em plus 0.5em
  minus 0.4em\relax IEEE, 2020, pp. 10\,402--10\,407.

\bibitem{2019szoke}
I.~Szöke, M.~Skácel, L.~Mošner, J.~Paliesek, and J.~Černocký, ``Building
  and evaluation of a real room impulse response dataset,'' \emph{IEEE J. of
  Sel. Topics in Signal Proc.}, vol.~13, no.~4, pp. 863--876, 2019, {D}ata:
  \url{https://speech.fit.vutbr.cz/software/but-speech-fit-reverb-database}.

\bibitem{McKenzie2021AcousticAA}
T.~McKenzie, S.~J. Schlecht, and V.~Pulkki, ``Acoustic analysis and dataset of
  transitions between coupled rooms,'' \emph{{ICASSP}}, pp. 481--485, 2021,
  {D}ata: \url{https://zenodo.org/record/4636068#.YhT-45PMJ6c}.

\bibitem{McKenzie2021DatasetOS}
T.~McKenzie, L.~McCormack, and C.~Hold, ``Dataset of spatial room impulse
  responses in a variable acoustics room for six degrees-of-freedom rendering
  and analysis,'' \emph{ArXiv}, vol. abs/2111.11882, 2021, {D}ata:
  \url{https://zenodo.org/record/5720724#.YhT_6JPMJ6c}.

\bibitem{Eaton2016EstimationOR}
J.~Eaton, N.~D. Gaubitch, A.~H. Moore, and P.~A. Naylor, ``Estimation of room
  acoustic parameters: The {ACE} challenge,'' \emph{IEEE/ACM Trans. on Audio,
  Speech, and Lang. Proc.}, vol.~24, pp. 1681--1693, 2016, {D}ata:
  https://doi.org/10.5281/zenodo.6257551.

\bibitem{di2021dechorate}
D.~Di~Carlo, P.~Tandeitnik, C.~Foy, A.~Deleforge, N.~Bertin, and S.~Gannot,
  ``d{E}chorate: a calibrated room impulse response database for echo-aware
  signal processing,'' \emph{arXiv preprint arXiv:2104.13168}, 2021,
  \url{https://doi.org/10.5281/zenodo.4626589}.

\bibitem{2017lippi}
M.~Lippi, M.~Mamei, S.~Mariani, and F.~Zambonelli, ``Coordinating distributed
  speaking objects,'' in \emph{Int. Conf. on Dist. Comp. Sys.}, 2017, pp.
  1949--1960.

\bibitem{2015mroueh}
Y.~Mroueh, E.~Marcheret, and V.~Goel, ``Deep multimodal learning for
  audio-visual speech recognition,'' in \emph{{ICASSP}}, 2015, pp. 2130--2134.

\end{thebibliography}
\end{document}